\documentclass[12pt]{article}
\usepackage{graphicx}
\usepackage{epsfig}
\begin{document}

\title{Ladder operators and coherent states for the trigonometric
  P\"oschl-Teller potential}
\author{R. Rom\'an-Ancheyta, $^{(1)}$ O. de los
  Santos-S\'anchez,$^{(1,2)}$ J. R\'ecamier$^{(1)}$ \\
$^{(1)}$ Instituto de Ciencias F\'{\i}sicas\\
 Universidad Nacional Aut\'onoma de M\'exico\\
 Apdo. Postal 48-3, Cuernavaca,
  Morelos 62251, M\'exico\\
 $^{(2)}$ Instituto de F\'{\i}sica\\
Benem\'erita Universidad Aut\'onoma de Puebla\\
 Apdo. Postal J-48, Puebla, Pue., M\'exico}
\date{14 March 2011}

\maketitle

\begin{abstract}
In this work we make use of {\em deformed} operators to construct the
coherent states of some nonlinear systems by generalization of two
definitions: {\em i)} As eigenstates of a deformed annihilation
operator and {\em ii)} by application of a deformed displacement
operator to the vacuum state. We also construct the coherent states
for the same systems using the ladder operators obtained by
traditional methods with the knowledge of the eigenfunctions and
eigenvalues of the corresponding Schr\"odinger equation. We show that
both methods yield coherent states with identycal algebraic structure.
\end{abstract}
\section{Introduction}
In this work we make use of two different methods to construct
ladder operators for different potentials and exemplify the
methodology with a well known example. On the one hand, we 
make use of the knowledge of the eigenfunctions of the corresponding
Schr\"odinger equation and apply a standard procedure for their
construction \cite{renato,chino}. On the other hand, we make use of
the idea of {\em deformed} oscillators introduced by Man'ko
\cite{manko1} and, by making the appropriate choice of the deformation
function, a Hamiltonian of the harmonic oscillator form written in
terms of deformed operators yields the eigenvalues of the
corresponding Schr\"odinger equation. Once we have constructed the
ladder operators with each one of these methodologies, we construct
their coherent states as eigenstates of the annihilation operator and
by the application of the displacement operator to the vacuum state.  
\section{Ladder operators}
In this section we use a standard procedure for the construction of
ladder operators for a system whose analytical solution is known. As an
 example we consider a trigonometric potential given by: 
\begin{equation}
V(x) = U_0 \tan^2(ax)
\end{equation}
where $U_0$ is the potential's strenght and $a$ its range. The number
of bound states for this potential is infinite. It's eigenfunctions
and eigenvalues are \cite{fonieto}:  
\begin{equation}
\psi_n^{\lambda}(x)=\sqrt{\frac{a(\lambda
    +n)\Gamma(2\lambda+n)}{\Gamma(n+1)}}\left(\cos(ax)\right)^{1/2}
P_{n+\lambda-1/2}^{1/2-\lambda}(\sin(ax)),  
\end{equation}
\begin{equation}\label{eq:espectro1}
E_n = \frac{\hbar^2a^2}{2\mu}(n^2+2n\lambda+\lambda) =\hbar
\omega(n+\frac{1}{2}+\frac{n^2}{2\lambda}), 
\end{equation}
where $\mu$ is the mass of the particle, $\omega =\hbar \lambda a^2/\mu$ and
the parameter $\lambda$ is related to the potential strength and range by 
$\lambda(\lambda+1)=2\mu U_0/\hbar^2 a^2$.\\
In the harmonic limit $\lambda \rightarrow \infty$ and
$a\rightarrow 0$ with $\lambda a^2 = \mu \omega/\hbar$. 

Making the change of variable $u = \sin(ax)$ we can write the
eigenfunctions as 
\begin{equation}
\psi_n^{\lambda}(u)=\sqrt{\frac{a(\lambda
    +n)\Gamma(2\lambda+n)}{\Gamma(n+1)}}\left(1-u^2\right)^{1/4}
P_{n+\lambda-1/2}^{1/2-\lambda}(u).  
\end{equation}
The action of the differential operator $d/du$ acting upon the
eigenfunctions $\psi_n^{\lambda}(u)$ can be obtained from its action
upon the Legendre polynomials:
\begin{equation}
(1-x^2)\frac{d P_{\nu}^{\mu}(x)}{dx} = (1+\nu)xP_{\nu}^{\mu}(x) -
  (\nu-\mu+1)P_{\nu+1}^{\mu}(x) 
\end{equation} 
 Calling 
 \[ N_{n}^{\lambda} = \sqrt{\frac{a(\lambda
    +n)\Gamma(2\lambda+n)}{\Gamma(n+1)}} =\sqrt{\frac{a(\lambda
    +n)\Gamma(2\lambda+n)}{n!}}
\] we get:
\begin{equation}
\frac{d \psi_{n}^{\lambda}(u)}{du} =
\left(\frac{n+\lambda}{1-u^2}u\right)\psi_{n}^{\lambda}(u) 
     -\frac{2\lambda+n}{1-u^2}\frac{N_{n}^{\lambda}} 
     {N_{n+1}^{\lambda}} \psi_{n+1}^{\lambda}(u)
\end{equation}
 rearranging terms, substituting the normalization constant and defining
$\epsilon=\lambda+n$ we obtain the following relation between the
eigenfunctions $\psi_n^{\lambda}(u)$ and $\psi_{n+1}^{\lambda}(u)$: 
\begin{equation}
(1-u^2)\left(-\frac{d}{du}+\frac{\epsilon}{1-u^2}u\right)
  \sqrt{\frac{\epsilon+1}{\epsilon}}\psi_{n}^{\lambda}(u) =
  \sqrt{(n+1)(2\lambda+n)}\psi_{n+1}^{\lambda}(u)  
\end{equation}
so that, the operator $M_{+}$ defined by
\begin{equation}
M_{+}=(1-u^2)\left(-\frac{d}{du}+\frac{\epsilon}{1-u^2}u\right)
\sqrt{\frac{\epsilon+1}{\epsilon}}   
\end{equation}
acting upon the function $\psi_{n}^{\lambda}(u)$ yields
\begin{equation}
M_{+}\psi_{n}^{\lambda}(u) = m_{+}\psi_{n+1}^{\lambda}(u),
\ \ \ m_{+}=\sqrt{(n+1)(2\lambda+n)}.
\end{equation}
In order to construct the annihilation operator $M_{-}$ we first make
use of the relationship \cite{tablas} 
\[ (1-x^2)\frac{dP_{\nu}^{\mu}(x)}{dx} = -\nu x P_{\nu}^{\mu}(x) +
(\nu+\mu)P_{\nu-1}^{\mu}(x). \]
Acting upon the eigenfunctions $\psi_{n}^{\lambda}(u)$ we
obtain
\begin{equation}
\frac{d \psi_{n}^{\lambda}(u)}{du} =-\frac{(n+\lambda)}{1-u^2}u
\psi_{n}^{\lambda}(u) + \frac{n}{1-u^2}
\frac{N_{n}^{\lambda}}{N_{n-1}^{\lambda}} \psi_{n-1}^{\lambda}(u)
\end{equation}
rearranging terms and substituting the explicit form of the
normalization constants we get:
\begin{equation}
(1-u^2)\left(\frac{d}{du}
+\frac{n+\lambda}{1-u^2}u\right)\sqrt{\frac{\lambda +n-1}{\lambda+n}}
\psi_{n}^{\lambda}(u) = \sqrt{n(2\lambda+n-1)} \psi_{n-1}^{\lambda}(u) 
\end{equation}
and we can define the annihilation operator 
\begin{equation}
M_{-} = (1-u^2)\left(\frac{d}{du} +
\frac{u}{1-u^2}\epsilon\right)\sqrt{\frac{\epsilon-1}{\epsilon}} 
\end{equation}
whose effect acting upon the wavefunction $\psi_{n}^{\lambda}(u)$ is
\begin{equation}
M_{-}\psi_{n}^{\lambda}(u) = m_{-}\psi_{n-1}^{\lambda}(u),
\ \ \ m_{-}=\sqrt{n(2\lambda +n -1)}
\end{equation} 
In order to recover the harmonic limit we define the operators $\hat
b^{\dagger}$, $\hat b$ as: 
\begin{equation}
\hat b^{\dagger} = \frac{1}{\sqrt{2\lambda}} M_{+}, \ \ \ \hat b =
\frac{1}{\sqrt{2\lambda}} M_{-}  
\end{equation}
whose action upon the eigenfunctions $\psi_{n}^{\lambda}$ is:
\begin{equation}\label{eq:accion1}
\hat b \psi_{n}^{\lambda} =
\sqrt{\frac{n(2\lambda+n-1)}{2\lambda}}\psi_{n-1}^{\lambda} =
b_{-}\psi_{n-1}^{\lambda} 
\end{equation}
and
\begin{equation}\label{eq:accion2}
\hat b^{\dagger} \psi_{n}^{\lambda} =
\sqrt{\frac{(n+1)(2\lambda+n)}{2\lambda}}\psi_{n+1}^{\lambda} =
b_{+}\psi_{n+1}^{\lambda} .
\end{equation}

The commutation relations between the operators $\hat b^{\dagger}$,
$\hat b$ are obtained from their action upon the eigenfunctions
$\psi_{n}^{\lambda}$:  
\begin{equation}
[\hat b^{\dagger}, \hat b]\psi_{n}^{\lambda} =\left(
\frac{n(2\lambda+n-1)}{2\lambda} - 
\frac{(n+1)(2\lambda+n)}{2\lambda}\right) \psi_{n}^{\lambda} =
-\frac{2(\lambda + n)}{2\lambda}\psi_{n}^{\lambda} 
\end{equation}
calling the operator
\[ \hat b_0 =  \frac{(\lambda + \hat n)}{\lambda} = 1 + \frac{\hat
  n}{\lambda} \] we obtain the commutation 
relations: 
\begin{equation}
[\hat b,\hat b^{\dagger}] = \
\hat b_0, \ \ \ [\hat b,\hat b_{0}] = \frac{1}{\lambda} \hat
b,\ \ \ [\hat b^{\dagger},\hat b_0] = -\frac{1}{\lambda}\hat
b^{\dagger}. 
\end{equation}
In the harmonic limit the operators $\hat b^{\dagger}$, $\hat b$ go into
\begin{equation}
\hat b^{\dagger} \rightarrow
-\sqrt{\frac{\hbar}{2\mu\omega}}\frac{d}{dx} +
\sqrt{\frac{\mu\omega}{2\hbar}} x 
\end{equation}
and
\begin{equation}
\hat b \rightarrow \sqrt{\frac{\hbar}{2\mu\omega}}\frac{d}{dx} +
\sqrt{\frac{\mu\omega}{2\hbar}} x 
\end{equation}
as they should.
\section{Deformed Oscillator}
In this section we introduce {\em deformed} boson creation and
annihilation operators $\hat A^{\dagger}$, $\hat A$ which differ from
the usual harmonic oscillator operators 
$\hat a$, $\hat a^{\dagger}$ by a deformation function of the number
operator, that is, 
\begin{equation}
\hat A = \hat a f(\hat n) = f(\hat n+1)\hat a, \ \ \ \hat A^{\dagger}
=f(\hat n) \hat a^{\dagger} = \hat a^{\dagger}f(\hat n+1).
\end{equation}
In this work we will consider real and positive functions of the
number operator $\hat n = \hat a^{\dagger} \hat a$.  \\
A Hamiltonian of the Harmonic oscillator form written in terms of the
deformed operators becomes
\begin{equation}\label{eq:Hdef1}
H_D = \frac{\hbar \Omega}{2}\left(\hat A^{\dagger}\hat A + \hat A \hat
A^{\dagger}\right) =  \frac{\hbar \Omega}{2}\left(\hat n f^2(\hat n)+
(\hat n+1)f^2(\hat n+1)\right).
\end{equation}
The explicit form of the energy spectra can be fixed when one
specifies the deformation function. As a first example let
us consider the trigonometric P\"oschl-Teller energy spectra given by
Eq.~\ref{eq:espectro1}. If we choose the deformation function as:
\begin{equation}\label{eq:defunct}
f^2(\hat n) = \frac{\hbar a^2}{2\mu\Omega}\left(\hat n+2\lambda
-1\right)
\end{equation}
the deformed Hamiltonian given by Eq.~\ref{eq:Hdef1} becomes
\begin{equation}
H_D = \frac{\hbar^2 a^2}{2\mu}\left(\hat n^2 +2\lambda \hat
n+\lambda\right)
\end{equation}
whose eigenvalues are identical with those of the  trigonometric
P\"oschl-Teller potential given in Eq.~\ref{eq:espectro1}.\\
The action of these operators over the states of number
$\vert n \rangle$ give as result
\begin{equation}
\hat A \vert n \rangle=
\sqrt{\frac{n(2\lambda+n-1)}{2\lambda_0}}\vert n-1 \rangle,
 \\\ \hat A^{\dagger}\vert n \rangle=
\sqrt{\frac{(n+1)(2\lambda+n)}{2\lambda_0}}\vert n+1 \rangle
\end{equation}
notice the similarity with Eqs.~\ref{eq:accion1} and
\ref{eq:accion2}. Because of this, if we take a hamiltonian of
the harmonic oscillator form in terms of the operators
$\hat b, \hat b^{\dagger}$ the eigenvalues will be the same as 
those shown in Eq.~\ref{eq:espectro1} i.e., they are isospectral.\\

Once we have fixed the deformation function, the operators
$\hat A$, $\hat A^{\dagger}$ are determined and the algebraic
properties of the system can be explored.
The commutation relations between the deformed operators are
\begin{equation}\label{eq:conmuta}
[ \hat A, \hat A^{\dagger}] = \frac{\hbar a^2}{\mu\Omega}\left(\hat n +
\lambda\right), \ \ \ [ \hat A, \hat n] = \hat A, \ \ \ [ \hat
    A^{\dagger},\hat n] = -\hat A^{\dagger},
\end{equation}
 notice that the set of operators $\{\hat A, \hat A^{\dagger}, \hat n,
 1\} $ is closed under the operation of commutation.
\section{Coherent states}
The first proposal for the construction of what is now
known as a coherent state was done by Schr\"odinger in 1926
\cite{schroedinger} in connection with the classical states of the
quantum harmonic oscillator. Much later, in 1963
Glauber\cite{glauber1} constructed the eigenstates of the annihilation
operator of the harmonic oscillator in order to study the
electromagnetic correlation functions and showed that such states are
enormously useful for quantum optics. At about the same
time, Klauder \cite{klauder} 
developed a set of continuous states in which the basic ideas of
coherent states for arbitrary Lie groups were contained. In the early
70's the complete construction of coherent states of Lie groups was
achieved by Perelomov \cite{perolomov} and Gilmore
\cite{gilmore}. The basic theme of this development was to connect the
coherent states with the dynamical group for each physical problem.
Since then, the idea of generalizing the concept of coherent states to
arbitrary systems has been considered, one of the difficulties being that the
different generalizations lead to different results
\cite{gilmore2,gazeau,reca08}. 
 For the harmonic oscillator
there are three alternative definitions for the construction of its
coherent states: {\em i)} as those states that saturate the dispersion
relations $\Delta x \Delta p = \hbar/2$, {\em ii)} as eigenstates of
the annihilation operator $\hat a \vert \alpha \rangle = \alpha \vert
\alpha \rangle$ and {\em iii)} by displacement of the vacuum state
$D(\alpha )\vert 0 \rangle = \vert \alpha \rangle$. In this work we
consider the generalization of two of the abovementioned definitions.    
\subsection{Coherent states as eigenstates of the annihilation
  operator}\label{seclop} 
Coherent states for the trigonometric P\"oschl-Teller potential can be
obtained as eigenstates of the annihilation operator $\hat b$  
\begin{equation}
\hat b \vert \alpha,\lambda \rangle = \alpha \vert \alpha,\lambda
\rangle 
\end{equation}
The states $\vert n,\lambda\rangle$ form a complete set, then we can
write the coherent state $\vert \alpha,\lambda\rangle$ as: 
\begin{equation}\label{eq:CSTP}
\vert \alpha,\lambda\rangle = \sum_{n=0}^{\infty} C_{n} \vert
n,\lambda\rangle 
\end{equation}
We have then, by application of the annihilation operator $\hat b$ to
the coherent state 
\begin{equation}
\hat b \vert \alpha, \lambda \rangle = \sum_{n=1}^{\infty} C_{n}
\sqrt{n(1+\frac{n-1}{2\lambda})} \vert n-1,\lambda\rangle = \alpha
\sum_{n=1}^{\infty} C_{n-1} \vert n-1,\lambda\rangle 
\end{equation}
and we thus find the relationship between the coefficients $C_{n}$ and
$C_{n-1}$ 
\[ C_n \sqrt{n\left(1+\frac{n-1}{2\lambda}\right)} = \alpha C_{n-1}. \]
Further applications of the annihilation operator yield relations
between $C_{n}$ and $C_{n-2}$, $C_{n-3}$,\dots until after $n$
applications we finally get a relationship between $C_n$ and $C_0$: 
\begin{equation}
C_n \sqrt{n(n-1)\cdots (1)\frac{(2\lambda +n -1)(2\lambda +n -2)\cdots
    (2\lambda)}{(2\lambda)^n}} = \alpha^n C_0. 
\end{equation}
Then, substituting into Eq.~\ref{eq:CSTP} the coherent state $\vert
\alpha, \lambda\rangle$ can be written as: 
\begin{equation}
\vert \alpha,\lambda \rangle = C_0 \sum_{n=0}^{\infty} \alpha^n
\sqrt{\frac{(2\lambda)^n \Gamma(2\lambda)}{n!\Gamma(2\lambda +n)}}
\vert n,\lambda\rangle 
\end{equation}
with $C_0$ a normalization constant. 
\subsection{Displacement Operator Coherent States}\label{displacement}
In this section we construct the coherent states by application of the
generalized displacement operator upon the vacuum state. In order to
do that it is necessary to make use of the commutation relations
between the ladder operators and the number operator. With the ladder 
operators we constructed in section 1 we have the commutation
relations:
\begin{equation}
[\hat b, \hat b^{\dagger}] = \hat b_0 = 1 + \frac{\hat n}{\lambda},
\ \ \ [\hat b, \hat b_0] = \frac{1}{\lambda}\hat b, \ \ \ [\hat
  b^{\dagger}, \hat b_0] = -\frac{1}{\lambda}\hat b^{\dagger},
\end{equation}
since the set of operators $\{\hat b, \hat b^\dagger, \hat b_0\}$ is
closed under commutation we can write:
\begin{equation}
D_D(\alpha)=\exp[\alpha \hat b^{\dagger} -\alpha^{*}\hat b]=
e^{\alpha_{+}\hat b^{\dagger}} e^{\alpha_0 \hat b_0} e^{\alpha_{-}\hat b}
\end{equation}
with complex functions $\alpha_{+}$, $\alpha_0$, $\alpha_{-}$ to be
determined. Using the results of Ref.~\cite{santos} we obtain:
\begin{eqnarray*}
D_D(\alpha)= \exp\left[ \frac{\alpha \tanh(\vert
    \alpha\vert/\sqrt{2\lambda})}{\vert \alpha \vert/\sqrt{2\lambda}}
  \hat b^{\dagger}\right] \exp\left[ -2\lambda \ln(\cosh(\vert
  \alpha\vert/\sqrt{2\lambda}))\hat b_0\right] \times \\ 
\exp\left[\frac{-\alpha^{*} \tanh(\vert \alpha
    \vert/\sqrt{2\lambda})}{\vert \alpha \vert /\sqrt{2\lambda}} \hat
  b\right] 
\end{eqnarray*}
with $\alpha = \vert \alpha \vert e^{i\phi}$ and $\zeta=
e^{i\phi}\tanh(\vert \alpha \vert /\sqrt{2\lambda})$, the deformed
displacement operator can be written as: 
\begin{equation}
D_D(\zeta) = \exp\left[\zeta \sqrt{2\lambda} \hat
  b^{\dagger}\right] (1-\vert \zeta\vert^2)^{\lambda \hat b_0}
\exp\left[ -\zeta^{*}\sqrt{2\lambda}\hat b\right] 
\end{equation}
Applying it to the vacuum state we obtain the coherent states
\begin{equation}
\vert \zeta \rangle =D_D(\zeta)\vert 0, \lambda \rangle = (1-\vert
\zeta\vert^2)^{\lambda} \sum_{n=0}^{\infty}
\sqrt{\frac{\Gamma(n+2\lambda)}{n!\Gamma(2\lambda)}}
\zeta^n \vert n, \lambda \rangle 
\end{equation}

\subsection{Coherent states as eigenstates of the deformed
  annihilation operator}\label{defanil}
 Let us now construct
their coherent states as eigenstates of the deformed annihilation
operator \cite{manko,santos}
\begin{equation}\label{eq:ecohda}
\hat A \vert \alpha,f\rangle = \alpha\vert \alpha,f\rangle
\end{equation}
As in section \ref{seclop} we express them as an expansion in the
basis of number states $\{ \vert 0 \rangle, \vert 1 \rangle, \dots,
\vert n \rangle, \dots \}$
\begin{equation}
 \vert \alpha, f\rangle =N_f \sum_{n=0}^{\infty} c_{n}^{f} \vert n \rangle, 
\end{equation}
substitution into Eq.~\ref{eq:ecohda} yields the following relation
between the coefficients $c_{n}^{f}$ and $c_{n-1}^{f}$:
\begin{equation}
c_n^f f(n)\sqrt{n} = \alpha c_{n-1}^f
\end{equation}
applying the annihilation operator $n$ times we obtain:
\begin{equation}
c_{n}^f f(n)!\sqrt{n!} = \alpha^n c_{0}^f
\end{equation}
where $f(n)! = f(n)f(n-1)\cdots f(0)$.\\
Then the coherent state is given by:
\begin{equation}\label{eq:aocs}
\vert \alpha,f\rangle = N_{f}\sum_{n=0}^{\infty}
\frac{\alpha^n}{\sqrt{n!}f(n)!} \vert n \rangle.
\end{equation}
Substitution of the explicit form of the deformation function yields
\begin{equation}
\vert \alpha,f\rangle = N_f \sum_{n=0}^{\infty}
\sqrt{\frac{(2\lambda_0)^{n}
    \Gamma(2\lambda)}{n!\Gamma(2\lambda + n)}} \alpha^n
\vert n \rangle
\end{equation}
with $N_f$ a normalization constant. Notice that the coherent states
we have constructed with the knowledge of the deformation function
have an identycal structure as those obtained via the ladder operators
in section \ref{seclop}.   

\subsection{Coherent states obtained via the deformed displacement
  operator}
From the commutation relations given by Eqs.\ref{eq:conmuta} we notice
that the set of operators $\{ \hat A, \hat A^{\dagger}, \hat n, 1\}$
constitute a finite Lie algebra. For convenience we define
\[ g(\hat n; a,\lambda)= \hbar a^2(\lambda + \hat n)/\mu \Omega=
\frac{\lambda +\hat n}{\lambda_0} \] with $\lambda_0=\mu\Omega/\hbar
a^2$ so
that the commutation relations given in Eqs.~\ref{eq:conmuta} become
\begin{equation}
[\hat A,  g(\hat n; a,\lambda)]=\frac{\hat A}{\lambda_0},
\ \ \ [\hat A^{\dagger},  g(\hat n; a,\lambda)]=-\frac{\hat
  A^{\dagger}}{\lambda_0}, \ \ \ [\hat A, \hat A^{\dagger}] =
g(\hat n; a,\lambda). 
\end{equation}
clearly, the set of operators $\{ \hat A, \hat A^{\dagger}, g(\hat n;
a,\lambda)\}$ is closed under commutation and the deformed displacement
operator $D_D(\alpha)$ obtained by the replacement of the usual
harmonic oscillator operators $\hat a$, $\hat a^{\dagger}$ by their
deformed counterparts can be expressed in terms of a product of
exponentials. That is, 
\begin{eqnarray}
D_D(\alpha) & = & \exp\left[ \alpha \frac{\tanh(\frac{\vert
      \alpha\vert}{\sqrt{2\lambda_0}})}{\frac{\vert \alpha
        \vert}{\sqrt{2\lambda_0}}} \hat A^{\dagger}  \right]  
\exp\left[-2\lambda_0\ln ( \cosh( \frac{\vert \alpha\vert}{\sqrt{2\lambda_0}}))
g(\hat n;a,\lambda)\right]\nonumber \\
& \times & \exp\left[ -\alpha^{*}\frac{\tanh(\frac{\vert
      \alpha\vert}{\sqrt{2\lambda_0}})}{\frac{\vert \alpha
      \vert}{\sqrt{2\lambda_0}}} \hat A  \right]. 
\end{eqnarray}
if we now define $\zeta_0 = e^{i\phi_0}\tanh(\vert
\alpha\vert/\sqrt{2\lambda_0})$ and write $\alpha = \vert \alpha \vert
e^{i\phi_0}$ we obtain after a little algebra:
\begin{equation}
D_D(\zeta_0)= \exp\left[\zeta_0 \sqrt{2\lambda_0} \hat A^{\dagger}\right]
\left( 1 -\vert \zeta_0 \vert^2\right)^{\lambda_0 g(\hat n;a,\lambda)}
\exp\left[-\zeta_0^{*} \sqrt{2\lambda_0} \hat A\right].
\end{equation}
 The coherent state obtained by application of this deformed
 displacement operator upon the vacuum state yields
\begin{equation}
\vert \zeta_0 \rangle = D_D(\zeta_0)\vert 0 \rangle = (1-\vert \zeta_0
\vert^2)^{\lambda}
\sum_{n=0}^{\infty}(2\lambda_0)^{n/2}\frac{\zeta_0^n}{\sqrt{n!}}
f(n)!\vert n \rangle
\end{equation}
replacing the explicit form of the deformation function we obtain
finally
\begin{equation}
\vert \zeta_0 \rangle =  (1-\vert \zeta_0 \vert^2)^{\lambda}
\sum_{n=0}^{\infty}
\sqrt{\frac{\Gamma(n+2\lambda)}{n!\Gamma(2\lambda)}} \zeta_0^n   \vert
n \rangle.  
\end{equation}
These coherent states have exactly the same algebraic structure as
those obtained with the  ladder operators $ \hat b$, $ \hat b^{\dagger}$
constructed in section \ref{displacement}. 
\subsection{Other examples}
We now consider a harmonic oscillator with an inverse square potential in
two dimensions. The potential has the form
\[ V(r) = \frac{1}{2}\mu \omega^2 r^2 +
\frac{\hbar^2}{2\mu}\frac{\alpha}{r^2} \] where $\omega$ is the
frequency of the oscillator and $\alpha$ is the potential
strength. As shown in  \cite{chinoijmpa}, the eigenfunctions are
\[ R_n(\rho)= N_n \rho^s e^{-\frac{\rho}{2}}L_n^{2s}(\rho), \ \ \ N_n
= \sqrt{\frac{2n!}{\Gamma(n+2s+1)}} \] and the eigenvalues are
\[ E_n = 2(n + s + \frac{1}{2}), \]
 where
$s=\sqrt{\alpha+m^2}/2$ with $m=0, 1, 2, \dots$, $n=0, 1, 2, \dots$,
 $L_n^{2s}(\rho)$ is the associated Laguerre Polynomial
and the units used were $\hbar=\mu=\omega=1$. The ladder operators
obtained after the application of the usual procedure are:
\begin{equation}
\hat {\cal L}_{-} = -\rho \frac{d}{d\rho} + s + \hat n
-\frac{\rho}{2}, \ \ \hat {\cal L}_{+} = \rho \frac{d}{d\rho} + s +
\hat n + 1 -\frac{\rho}{2}
\end{equation}  
whose action upon the eigenfunctions is:
\begin{equation}
\hat{\cal L}_{-}\vert n, \rho\rangle = \sqrt{n(n+2s)}\vert
n-1,\rho\rangle, \ \ \hat{\cal L}_{+}\vert n, \rho\rangle =
\sqrt{(n+1)(n+2s+1)}\vert n+1,\rho\rangle.
\end{equation}
and their commutation relations are
\begin{equation}
[ \hat{\cal L}_{-},\hat{\cal L}_{+}] = 2\hat{\cal L}_0,\ \  [\hat{\cal
    L}_0, \hat{\cal L}_{-}]=-\hat{\cal L}_{-}, \ \ [\hat{\cal L}_0,
  \hat{\cal L}_{+}] = \hat{\cal L}_{+}. 
\end{equation}
where the operator $\hat{\cal L}_0 = \hat n +s +\frac{1}{2}$. 

Following the methodology of section \ref{seclop} we obtain the
coherent states:
\begin{eqnarray}
\vert \alpha \rangle = C_0 \sum_{n=0}^{\infty}
\frac{\alpha^n}{\sqrt{n!(2s+n)(2s+n-1)\cdots (2s+1)}} \vert n,
\rho\rangle \\ \nonumber = C_0 \sum_{n=0}^{\infty}\alpha^n
\sqrt{\frac{\Gamma(2s+1)}{n!\Gamma(2s+n+1)}} \vert n, \rho\rangle.
\end{eqnarray}
When using deformed operators for the trigonometric potential we found
that a deformation function of the form given by Eq.~\ref{eq:defunct}
could reproduce the energy spectra. In this case there is no quadratic
term in the spectra so we will take a antisymmetric combination for
the factorization of the Hamiltonian. That is:
\begin{equation}\label{eq:asymm}
H_D = \hat A \hat A^{\dagger} - \hat A^{\dagger} \hat A = (\hat
n+1)f^2(\hat n+1)-\hat nf^2(\hat n).
\end{equation}
 We propose a deformation function of the form
\begin{equation}
f^2(\hat n) = a\hat n + b
\end{equation}
with $a$ and $b$ real constants to be determined. Substitution into
Eq.~\ref{eq:asymm} and comparaison with the eigenvalues 
\[  (\hat
n+1)f^2(\hat n+1)-\hat nf^2(\hat n)=2(\hat n a + \frac{a+b}{2}) =
2(\hat n + s + \frac{1}{2})
\] yield $a=1$, $b=2s$ so that the deformation
function for this potential is:
\begin{equation}
f^2(\hat n) = \hat n + 2s.
\end{equation} 
Once we have selected the deformation function, the deformed
operators are specified. Their commutation relations are:
\begin{equation}
[\hat A, \hat A^{\dagger}] = 2(\hat n+s+\frac{1}{2}) = 2\hat A_0, \ \
[\hat A_0, \hat A] = -\hat A,\ \ \ [\hat A_0, \hat A^{\dagger}] = \hat A^{\dagger}. 
\end{equation}

The coherent states obtained as eigenstates of the annihilation
operator can be obtained by direct application of
Eq.~\ref{eq:aocs}. The result is: 
\begin{eqnarray}
\vert \alpha, f\rangle =
N_f\sum_{n=0}^{\infty}\frac{\alpha^n}{\sqrt{n!(n+2s)(n+2s-1)\cdots
    (2s+1)}}\vert n \rangle \\ \nonumber
=N_f \sum_{n=0}^{\infty} \alpha^n
\sqrt{\frac{\Gamma(2s+1)}{n!\Gamma(2s+n+1)}}\vert n \rangle
\end{eqnarray}
which have exactly the same structure as those obtained as eigenstates
of the annihilation operator $\hat {\cal L}_{-}$. 
Since the algebraic properties of the operators $\{ \hat{\cal L}_{+},
\hat{\cal L}_{-}, \hat{\cal L}_0 \}$ and the deformed operators $\{
\hat A, \hat A^{\dagger}, \hat A_0 \}$ are the same, the coherent
states obtained by application of the generalized displacement
operator with either set will have the same structure. 

\section{Discusion}
In this work we constructed the ladder operators corresponding to a
trigonometric potential and to a harmonic oscillator with an inverse
square potential and with them we found their coherent
states by generalization of two definitions. As 
eigenstates of the annihilation operator and as those states obtained
by the displacement of the vacuum state. We also made use of {\em
  deformed} operators with the deformation function chosen in
order to reproduce the energy spectra of each case. For the
trigonometric potential the deformation function was such that a
Hamiltonian of the harmonic oscillator form written 
in terms of these operators gave the same energy spectra. For the
other case we used a antisymmetric combination of the deformed
operators in order to reproduce the energy spectra. With these
operators we also constructed the coherent states using  the two
generalizations mentioned above. We found that the coherent states
obtained by the two methods have identycal algebraic structure.
 We have also considered other potentials whose eigenfunctions and
 eigenvalues are known and whose ladder operators have been
 obtained as for instance the Morse potential,
the hyperbolic P\"oschl-Teller potential and the ring-shaped non
spherical oscillator \cite{lemusbernal,chino,chino2}. From the
knowledge of the energy spectra we have obtained the deformation
function and the corresponding deformed operators. In all the above
mentioned cases the coherent states obtained with the deformed
 operators and those obtained with the ladder operators have identycal
 structure.\\ 
We have to mention that the deformed operators method, which we use for
obtain coherent states of the systems analyze, is much faster, powerful 
and elegant than the ladder operators method, constructed from the 
eigenfunctions of the corresponding Schrödinger equation. Due to the 
first one doesn't need to know the explicit form of these, just enough 
to know the energy spectra of the system and it comes down to find the 
deformation function that reproduce the spectrum.

\end{document}